\documentstyle[prl,aps,epsfig]{revtex}
\tighten
\begin{document}
\twocolumn

\title{Irrelevance of photon events distinguishability in a class of Bell
experiments}
\author{{Marek \.Zukowski$^1$, Dagomir Kaszlikowski$^1$ and 
Emilio Santos$^2$}\\
{\protect\small\sl $^1$Instytut Fizyki Teoretycznej i Astrofizyki,}\\
{\protect\small\sl Uniwersytet Gda\'nski, PL-80-952 Gda\'nsk, Poland}\\
{\protect\small\sl $^2$ 
Departamento de Fisica Moderna, Universidad de Cantabria, Santander, Spain }\\ 
}

\date{\today}
\maketitle

\begin{abstract}
We show that  the possibility 
of distinguishing between single and two photon detection events, usually not
met in the actual experiments, is not a necessary requirement for the proof 
that the 
experiments of Alley and Shih [Phys. Rev. Lett. {\bf 61}, 2921 (1988)], 
and Ou  and Mandel [Phys. Rev. Lett. {\bf 61}, 50 (1988)], are modulo fair
sampling assumption, valid tests of local realism.
We also give the critical parameters for the experiments to be
unconditional tests of local realism, and show that some other interesting
phenomena (involving bosonic type particle indistinguishability) can be 
observed during such tests.
\end{abstract}


The first Bell-type experiments which employed parametric 
down conversion process as the source of entangled photons
were those reported in refs \cite{SHIH} and  \cite{MANDEL}.
However, the specific traits of those experiments have led to a 
protracted dispute
on their validity as tests of local realism.
In this case the issue was not the standard problem of detection efficiency
(which up till now permits a local realistic interpretation of
all performed experiments).
The  trait that distinguishes the experiments is that even in the  
perfect {\it gedanken} situation (which assumes perfect detection) 
only in $50\%$ of the detection events 
each observer receives a
photon, in the other $50\%$ of events one observer receives both
photons of a pair while the other observer receives none.
The early ``pragmatic" approach was  to discuss only the events of the 
first type (as only such ones lead to spatially
separated coincidences). 
Only those were used as the data input to the
Bell inequalities in \cite{SHIH} and \cite{MANDEL}. 
This procedure was soon 
challenged (see e.g. \cite{KWIAT1} \cite{KWIAT2}, and 
especially in the theoretical analysis of ref. \cite{GARUCCIO}), as it 
raises justified doubts whether such
experiments could be ever genuine tests of local realism (as the effective
overall collection efficiency of the photon pairs, $50\%$ in the gedanken
case, is much below what is usually required for tests of local realism). 
Ten years after the first experiments  of this type were made, finally
the dispute was resolved \cite{SHZ}.
It was proposed, to take into account also those
``unfavourable" cases and to
analyse the entire
pattern of events. 
In this way one can indeed show that  the experiments 
are true test of local realism (namely, 
that the CHSH inequalities are violated by quantum predictions
for the idealised case). The idea was based upon
a specific value assignment for the ``wrong events" (see further, or
\cite{SHZ} itself). However, the scheme presented by 
Popescu et al \cite{SHZ} has one drawback. The authors assumed in 
their analysis that the detecting scheme employed in the 
experiment should be able to 
distinguish between two and one photon detections. 
This was not the case in the actual experiments.
The aim of this work is to show that even this is unnecessary, all one 
needs 
is the use of the specific value assignment procedure of \cite{SHZ}.

What is perhaps even more important, problems similar to those 
sketched above are also shared by the new, potentially highly
promising, class of EPR-Bell type experiments, which involve
the entanglement swapping procedure \cite{EVENT}. Also in this case the first
performed experiment did not employ detectors which are 
able to distinguish between firings caused by two photons and
a single photon \cite{BOUW}. The entanglement swapping
experiments thus far did not violate the visibility threshold
for local realism ($71\%$), however in the future the problem 
of their relation to the Bell theorem will be of a fundamental 
importance (as entanglement swapping may find application in future
quantum communication schemes \cite{BOSE}). 
The analysis presented in \cite{SHZ} can be adapted
to describe such experiments, clearly indicating violation of local realism.

Finally, we shall also give 
prediction of all effects occurring in the experiment.
It is quite often overlooked that a kind of Hong-Ou-Mandel dip phenomenon
\cite{HOM} can be observed in the experiment. 

In the class of experiments we consider
(fig.
1) \cite{SHZ} 
a type I parametric down-conversion source  \cite{MANDEL2} 
is used to generate pairs of
photons which are degenerated in frequency and  polarisation
(say $\hat x$) but propagate in two different directions. 
One of the photons
passes through a wave plate ($WP$) which rotates its polarisation by 
$90^o$.
The two photons are then directed 
onto the two input ports of a (nonpolarizing) ``$50-50$''
beamsplitter ($BS$). The observation stations
 are located in the exit beams of the beamsplitter. Each
local observer 
is equipped with a polarising beamsplitter \cite{PBS}, orientated along an
arbitrary axis (which, in principle can be 
randomly chosen, in the delayed-choice manner, 
just before the photons are supposed to
arrive). Behind each polarising beamsplitter are two detectors, $D_1^+$,
$D_1^-$ and $D_2^+$, $D_2^-$ respectively, where the lower index indicates
the corresponding observer and the upper index the two exit ports of the
polarised beamsplitter ($``+"$ meaning parallel with the polarisation axis of
the beamsplitter and $``-"$ meaning orthogonal to this axis). All
optical paths are assumed to be equal.

\begin{figure}[htbp]
     \begin{center}
       \includegraphics[angle=270, width=0.5\textwidth]{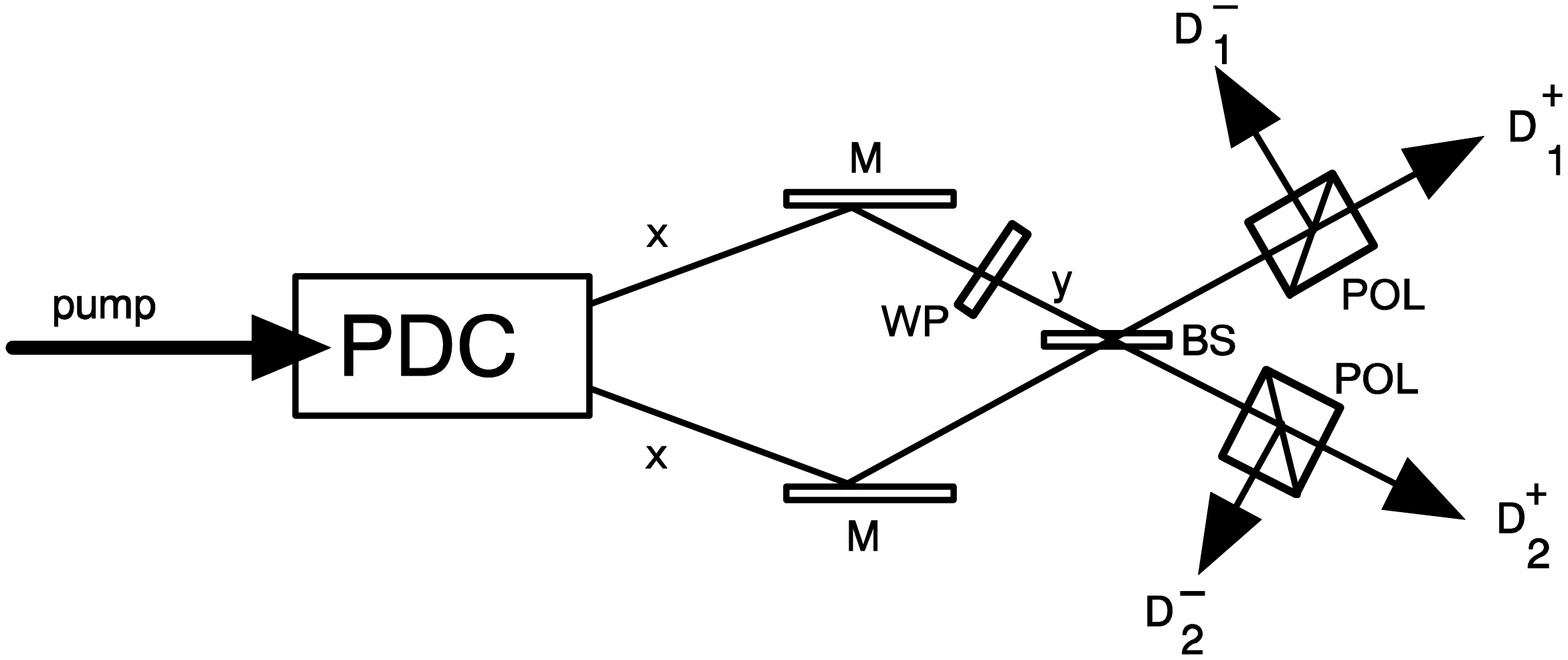}
       \caption{Schematic of the setup. For explanations see the 
main text}  \label{fig:det1}
     \end{center} 
   \end{figure}

Let us calculate the quantum predictions for the experiment.
We will use the second quantisation 
formalism, which is very convenient here, since the whole phenomenon observed
in the experiment rests upon the indistinguishability 
of photons. 

After the action of the wave-plate one can approximate quantum mechanical 
state describing two 
photons emerging from a non - linear crystal along the "signal" and the 
"idler"
beam by

\begin{equation}
|\Psi_{0}\rangle =a_{1\vec{x}}^{\dagger}a_{2\vec{y}}^{\dagger}|0\rangle,
\end{equation}
where $a_{1\vec{x}}^{\dagger}$ and $a_{2\vec{y}}^{\dagger}$ are creation
operators and $|0\rangle$ denotes the vacuum state. Subscripts 
$\vec{x},\vec{y}$
decode the  polarisation of the photon (either along $\vec{x}$ 
or $\vec{y}$ axis). The beamsplitter action can be described by

\begin{eqnarray}
&a_{1\vec{x}}^{\dagger}={1\over\sqrt{2}}(ic_{\vec{x}}^{\dagger}
+d_{\vec{x}}^{\dagger})
&\nonumber\\
&a_{2\vec{y}}^{\dagger}={1\over\sqrt{2}}(c_{\vec{y}}^{\dagger}+
id_{\vec{y}}^{\dagger}),&\\
\nonumber
\end{eqnarray}
where $c_{\vec{x}}^{\dagger},d_{\vec{x}}^{\dagger}
,c_{\vec{y}}^{\dagger},d_{\vec{y}}^{\dagger}$
are operators describing output modes of the beamsplitter ($c$ stands for
the first observer and $d$ for the second one). Thus
our state $|\Psi_{0}\rangle$ changes to :

\begin{equation}
|\Psi\rangle=
\frac{1}{2}(ic_{\vec{x}}^{\dagger}c_{\vec{y}}^{\dagger}-
c_{\vec{x}}^{\dagger}
d_{\vec{y}}^{\dagger}+c_{\vec{y}}^{\dagger}d_{\vec{x}}^{\dagger}+
id_{\vec{x}}^{\dagger}d_{\vec{y}}^{\dagger})|0\rangle
\end{equation}

Next comes the action of the polarisers in both beams, 
which can be described 
as

\begin{eqnarray}
&n_{\vec{x}}^{\dagger}=\cos(\theta_{1})n_{\parallel}^{\dagger}+
\sin(\theta_{1})n_{\perp}^{\dagger}& \nonumber\\
&
n_{\vec{y}}^{\dagger}=\sin(\theta_{1})n_{\parallel}^{\dagger}-
\cos(\theta_{1})n_{\perp}^{\dagger},& \nonumber\\
\end{eqnarray}
where $n^{\dagger}=c^{\dagger}$ or $d^{\dagger}$, and $n_{\parallel}^{\dagger}$ describes 
the mode parallel 
to
polarizer's axis  and $n_{\perp}^{\dagger}$ describes the mode perpendicular
to polarizer's axis; $\theta$ is the angle between the $\vec{x}$
axis and polarizer's axis. 
Thus the final state reaching the detector reads
\begin{eqnarray}
&|\psi_{final}\rangle=\frac{1}{2}\big[\sin(\theta_{1}-
\theta_{2})|c_{\parallel},
d_{\parallel}\rangle&\nonumber\\
&+\cos(\theta_{1}-\theta_{2})|c_{\parallel}
d_{\perp}\rangle&\nonumber\\
&
-\cos(\theta_{1}-\theta_{2})|c_{\perp},d_{\parallel}\rangle
+\sin(\theta_{1}-\theta_{2})|c_{\perp},d_{\perp}\rangle
&\nonumber\\
&+i{1\over{\sqrt{2}}}\sin(2\theta_{1})|2c_{\parallel}\rangle
+i{1\over{\sqrt{2}}}\sin(2\theta_{1})|2c_{\perp}\rangle
&\nonumber\\
&-i\cos(2\theta_{1})|c_{\perp},c_{\parallel}\rangle
+i{1\over{\sqrt{2}}}\sin(2\theta_{2})|2d_{\parallel}\rangle
&\nonumber\\
&+i{1\over{\sqrt{2}}}\sin(2\theta_{2})|2d_{\perp}\rangle
-i\cos(2\theta_{2})|d_{\parallel},d_{\perp}\rangle\big],&\nonumber\\
\label{final}
\end{eqnarray}
where e.g. $|c_{\parallel},
d_{\parallel}\rangle$ denotes one photon in the mode $c_{\parallel}$, and
one in
$d_{\parallel}$, whereas $|2c_{\parallel}\rangle={1\over{\sqrt{2}}}
{c_{\parallel}^{\dagger}}^2|0\rangle$ denotes two photons
in the mode $c_{\parallel}$.

Let us denote by $P(i,\theta_{1};j,\theta_{2})$ the joint
probability for the outcome $i$ to be registered by observer 1 when
her 
polariser is oriented along the direction that makes an angle $\theta_{1}$
with the $\vec{x}$ direction and the outcome $j$ to be registered by
observer 2 when her polariser is oriented along the direction that
makes an angle $\theta_{2}$ with the $\vec{x}$ direction. Here
$i,j=1-6$ and have the following meaning \cite{SHZ}:

1=one photon in $D^{-}$, no photon in $D^{+}$

2=one photon in $D^{+}$, no photon in $D^{-}$

3=no photons

4=one photon in $D^{+}$ and one photon in $D^{-}$

5=two photons in $D^{+}$, no photon in $D^{-}$

6=two photons in $D^{-}$, no photons in $D^{+}$.

The quantum predictions for joint probabilities of those events
are given by:

\begin{eqnarray}
&P(1,\theta_{1};1,\theta_{2})=P(2,\theta_{1};2,\theta_{2})={1\over 8}
[1-\cos2(\theta_{1}-\theta_{2})],&\\
&P(2,\theta_{1};1,\theta_{2})=P(1,\theta_{1};2,\theta_{2})={1\over 8}
[1+\cos2(\theta_{1}-\theta_{2})],&\\
&P(5,\theta_{1};3,\theta_{2})=P(6,\theta_{1};3,\theta_{2})={1\over{8}}
\sin^2(2\theta_{1}),&\\
&P(3,\theta_{1};5,\theta_{2})=P(3,\theta_{1};6,\theta_{2})={1\over{8}}
\sin^2(2\theta_{2}),&\\
&P(4,\theta_{1};3,\theta_{2})={1\over{4}}\cos^2(2\theta_{1}),&\\
&P(3,\theta_{1};4,\theta_{2})={1\over{4}}\cos^2(2\theta_{2}).&
\end{eqnarray}
Following \cite{SHZ} we associate with each outcome registered 
by the observer 
1 and 2
a corresponding value $a_{i}$ and
$b_{j}$ respectively, where $a_{1}=b_{1}=-1$
while all the other values are equal to 1. Let us denote by
$E(\theta_{1},\theta_{2})$ the expectation value of their product

\begin{equation}
E(\theta_{1},\theta_{2})
=\sum_{i,j}a_{i}b_{j}P(i,\theta_{1};j,\theta_{2}).
\end{equation}
After simple calculations one has:

\begin{eqnarray}
&E(\psi_{1},\psi_{2})&\nonumber\\
&=-{1\over{2}}\cos(\psi_{1}+\psi_{2})+\frac{1}{2},&
\label{perfect}
\end{eqnarray}
where we have put $2\theta_{k}=(-1)^{k-1}\psi_{k}$.

The above formula for the correlation function is valid if one 
assumes that it is possible to
distinguish between single and double photon detection. This is 
usually not
the case. Thus it is convenient to have a parameter $\alpha$ that
measures the distinguishability of the double and single detection at
one detector ( $0\leq\alpha\leq1$, and gives
the probability of distinguishing by the employed detecting scheme 
 of the double counts). The partial distinguishability blurs the 
distinction
between events 1 and 6 (2 and 5) and thus part of the events of
the type 6 are interpreted as of type 1 and are ascribed by the local
observer a wrong value, e.g. an event of type 6, if both photons
go to the $`` - "$ exit of the polariser, can be interpreted as a firing due 
to a single photon and is ascribed a $-1$ value. Please note that
such events like 1 or 2 in station 1 accompanied by 3 (no photon) at station 2
do not contribute to the correlation function because for any
$\alpha$ $P(1,\theta_{1};3,\theta_{2})=P(2,\theta_{1};3,\theta_{2})$.

If the parameter $\alpha$ is taken into account the correlation function 
acquires
the following form:

\begin{eqnarray}
&E(\psi_{1},\psi_{2}; \alpha)=&\nonumber\\
&=-{1\over{2}}\cos(\psi_{1}+\psi_{2})
+{1\over{2}}\alpha&\nonumber\\
&+{1\over{4}}(1-\alpha)(\cos^2\psi_{1}+\cos^2\psi_{2})&
\label{quant}
\end{eqnarray}
In this case after the insertion of the  quantum correlation
function (\ref{quant}) into the CHSH
inequality, 

\begin{eqnarray}
&-2\leq E(\psi_{1},\psi_{2};\alpha)+E(\psi_{1}',\psi_{2};\alpha)& \nonumber\\
&+E(\psi_{1},\psi_{2}';\alpha)-E(\psi_{1}',\psi_{2}';\alpha))\leq2,& \nonumber
\end{eqnarray}
one obtains:

\begin{eqnarray}
&-2\leq -{1\over{2}}[\cos(\psi_{1}+\psi_{2})+\cos(\psi_{1}'+\psi_{2})
&\nonumber\\
&+\cos(\psi_{1}+\psi_{2}')-\cos(\psi_{1}'+\psi_{2}')]+
\alpha &\nonumber\\
&+{1\over{2}}(1-\alpha)(\cos^2\psi_{1}+
\cos^2\psi_{2})\leq 2.&
\end{eqnarray}

The interesting feature of this inequality is that it can be violated for all
values of $\alpha$. What is perhaps even more important, it can be 
robustly violated even when one is not able to distinguish between
single and double clicks at all ($\alpha=0$). The actual value of
the CHSH expression can reach in this 
case $2.33712$ (a numerical result), 
which is only slightly less than the maximal value
for $\alpha=1$, which is $\sqrt{2}+1\approx2.41421$. Therefore we
conclude that in the experiment {\it one can observe violations of
local realism even if one is not able to distinguish between
the double and single counts at one detector}. That is, the essential
feature of the method of \cite{SHZ} to reveal
violations of local realism in the experiment of this type is the
specific value assignment scheme and not the double-single photon
counts distinguishability. 

The specific angles at which the maximum violation of 
the CHSH inequality is achieved for $\alpha=0$ differ very much from those 
for $\alpha=1$ (for which the standard result is reproduced), 
and they read (in radians) $\psi_1=2.93798$,
$\psi_1'=4.25513$, $\psi_2=-0.20241$ and $\psi_2'=1.11708$.

Let us notice that with the setup of 
fig.1 one is able to observe effects of similar
nature to the famous Hong-Ou-Mandel dip \cite{HOM}. These are 
revealed by the probabilities pertaining to the wrong events (8-11).
Simply for certain orientations of the polarisers, if the two photons emerge 
on 
one side of the experiment only, then they must exit the 
polarising beamsplitter via a single output port
(this effect is due to the bosonic-type indistinguishability of 
photons, see \cite{HOM}). 

Finally let us discuss what is the critical efficiency of the detection of the 
experiments of this type.
To this end, in our calculations we will use a very simple model of imperfect 
detections: 
we insert a beamsplitter with reflectivity $\sqrt{1-\eta}$,
in front of an ideal detector, which observes only the transmitted light. This 
results in the
system behaving just like a detector of efficiency $\eta$. If we assume that 
the incoming light
is described by a creation operator $a^ \dagger$ then transmitted
mode is denoted as $t_{a}^ \dagger$ whereas reflected
mode is denoted as $r_{a}^ \dagger$ and one has
:

\begin{equation}
a^ \dagger=\sqrt{1-\eta}r_{a^ \dagger}^ \dagger + \sqrt{\eta}t_{a^ \dagger}^ 
\dagger.
\end{equation}
For instance, if one takes the following part of the state vector 
(\ref{final}):
\begin{equation}
c_{||}^ \dagger d_{||}^ \dagger|0\rangle.
\end{equation}
the beamsplitter model of an imperfect
 detector transforms this term into:

\begin{eqnarray}
&[(1-\eta)r_{c_{||}}^ \dagger r_{d_{||}}^ \dagger +
\sqrt{\eta (1-\eta)}r_{c_{||}}^ \dagger t_{d_{||}}^ \dagger&\nonumber\\
& +
\sqrt{\eta (1-\eta)}t_{c_{||}}^ \dagger r_{d_{||}}^ \dagger +
\eta t_{c_{||}}^ \dagger t_{d_{||}}^ \dagger]|0\rangle.&\nonumber\\
\end{eqnarray}
The probabilities now read:

\begin{eqnarray}
&&P(3,\theta_{1};2,\theta_{2})=P(2,\theta_{1};3,\theta_{2}) \nonumber\\
&&P(1,\theta_{1};3,\theta_{2})=P(3,\theta_{1};1,\theta_{2})=
\eta (1-\eta)\\
&&P(1,\theta_{1};1,\theta_{2})=P(2,\theta_{1};2,\theta_{2})=
{1\over 4}\eta^2[\sin(\theta_{1}-\theta_{2})]^2\\
&&P(2,\theta_{1};1,\theta_{2})=P(1,\theta;2,\theta_{2})=
{1\over 4}\eta^2[\cos(\theta_{1}-\theta_{2})]^2\\
&&P(5,\theta_{1};3,\theta{2})=P(6,\theta_{1};3,\theta_{2})=
{1\over 8}\eta^2[\sin(2\theta_{1})]^2\\
&&P(3,\theta_{1};5,\theta{2})=P(3,\theta_{1};6,\theta_{2})=
{1\over 8}\eta^2[\sin(2\theta_{2})]^2\\
&&P(4,\theta_{1};3,\theta_{2})={1\over 4}\eta^2[\cos(2\theta_{1})]^2\\
&&P(3,\theta_{1};4,\theta_{2})={1\over 4}\eta^2[\cos(2\theta_{2})]^2
\end{eqnarray}

The correlation function, which includes the inefficiency of the detection 
reads

\begin{equation}
E(\psi_{1},\psi_{2};\eta,\alpha)=\eta^2E(\psi_{1},\psi_{2};\alpha)
+ (1-\eta)^2,
\end{equation}
where
$E(\psi_{1},\psi_{2};\alpha)$ is given by (\ref{quant}). 
We have tacitly assumed here that the parameters $\alpha$ and $\eta$
are independent of each other (this assumption may not hold for 
specific technical arrangements).
Putting this prediction into CHSH inequality, assuming that $\alpha=1$
(full distinguishability) we obtain a minimum
quantum efficiency needed for violation of local realism equal to $0.91$, 
whereas for other values of $\alpha$ we have: for $\alpha=0$ $\eta=0.926$;
for $\alpha=0.5$ $\eta=0.92$;
for $\alpha=0.75$ $\eta=0.92$;
for $\alpha=0.875$ $\eta=0.91$. One should note here that the method of value
assignment of \cite{SHZ} is in accordance with the method given by Garg and 
Mermin \cite{GM}
for the optimal estimation of required detector quantum efficiency to violate 
local realism
in a Bell-test. Thus the obtained efficiencies are indeed the lowest possible, 
and show that
experiments of this type are not good candidates for a "loophole-free" 
Bell-test \cite{SANTOS}, nevertheless due to the fact that the whole 
observable effect is a consequence of quantum principle of 
particle indistinguishability
such test are very interesting by themselves - they reveal the
entanglement inherently associated with this principle.

MZ was supported by the University of Gdansk Grant No
BW/5400-5-0202-8. DK was supported by the KBN Grant 2 P03B 096 15.


\begin{thebibliography}{99}
\bibitem{PBS} Following references \cite{GARUCCIO} and \cite{SHZ}
we assume that both local detection stations 
are equipped with polarising beamsplitters, and 
each of the output ports is observed by a detector. 
In the actual experiments \cite{SHIH} and \cite{MANDEL} at each station 
only one of the 
outputs was monitored.
\end{thebibliography}
\end{document}